\documentclass[preprint,showkeys,aps,prb]{revtex4}
\usepackage[dvips]{graphicx}

\begin{document}

\title{From Ann Arbor to Sheffield: Around the World in 80 Years. I.}

\author{
William Graham Hoover                           \\
Ruby Valley Research Institute                  \\
Highway Contract 60, Box 601                    \\
Ruby Valley, Nevada 89833                       \\
}

\date{\today}


\vspace{0.1cm}

\begin{abstract}
Childhood and graduate school at Ann Arbor Michigan prepared Bill for an interesting and rewarding career in physics.
Along the way came Carol and many joint discoveries with our many colleagues to whom we both owe this good life. This
summary of Bill's early work prior to their marriage and sabbatical in Japan is Part I, prepared for Bill's 80th
Birthday celebration at the University of Sheffield in July 2016.
\end{abstract}

\maketitle

\section{Bill's Early History in Chemistry and Physics, 1936-1990}

\subsection{ Ann Arbor Michigan to Washington D C to Oberlin Ohio and Back}

My research career got off to an early start in Ann Arbor. My earliest memories go back there to the late 1930s. My Father
taught econometrics and location theory at the University. The building always smelled of cigars.  In his lecture room he
had a giant wall-mounted sliderule and a desktop framework he had constructed with a few hundred vertical wires holding
just as many gold-colored cubes.  Each cube could slide up or down its $(x,y)$ support wire to represent an $(x,y,z)$
point on a surface in three-dimensional space.  Our Family moved to Washington D C during the war years for government work on
rationing and price controls. My Father, though assigned to the Navy, spent the last year of the war in Germany in a
Jeep, estimating the cost of rebuilding.

Back at home I was fascinated by mathematics, particularly the identity :
$$
\sum_1^N n^3 = \sum_1^N n \times \sum_1^N n \longrightarrow 1^3 + 2^3 + 3^3 + \dots = (1 + 2 + 3 + \dots )^2 \ .
$$
At Woodrow Wilson High School Louise ``Quiz'' Stull taught a rigorous and stimulating chemistry class. I was inspired by
her example and that of my chemist uncle John Redfield Hoover, a plastics enthusiast and fan of Dixieland Jazz. After
graduation I set off to Oberlin Ohio with a Procter and Gamble scholarship, planning to learn more chemistry. I
particularly enjoyed Luke Steiner's chemistry and thermodynamics courses there. Like Mrs. Stull Professor Steiner
used surprise quizzes to keep his class alert.  In his thermodynamics course we integrated heat capacity data by
counting squares on large-size graph paper.

\begin{figure}
\includegraphics[width=4.0in,angle=+90.]{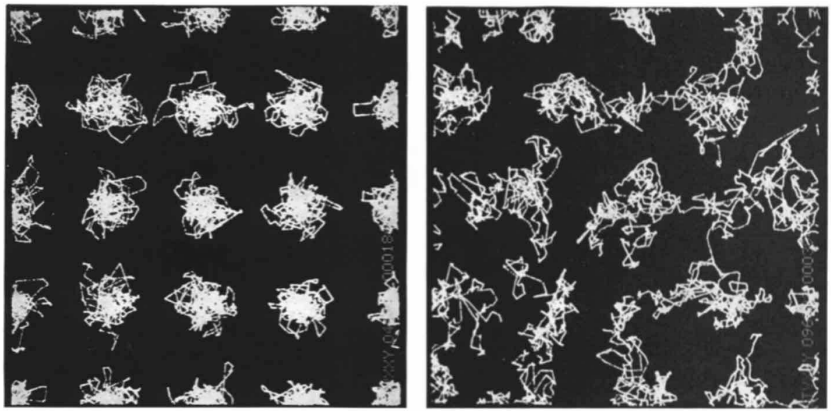}
\caption{
Trajectories for 32 hard spheres in the solid ( left ) and fluid ( right ) phases.
}
\end{figure}

In my third year at Oberlin I had an automobile accident and missed a
semester of school.  As a fringe benefit I chose to take Stuart Rice's course in statistical mechanics at Harvard
that summer, while earning a little money in my spare time helping measure cats' brain waves at M I T. After graduating
from Oberlin I returned to Michigan in 1958 for a PhD in chemical physics with Andy De Rocco, a gifted and fun-loving
teacher of statistical mechanics, with his course based on Joaquin Luttinger's lecture notes.  I also took Frank
Harary's graph-theory course in the mathematics department.  Graph theory came in handy years later in Livermore in
working with Francis Ree on the Mayers' virial series. We related the graphs for pressure with those for the pair
distribution function, and to the many-fewer but more-sophisticated graphs that have by now been successfully pursued
through the tenth virial coefficient for hard disks and hard spheres\cite{b1}.

There was a FORTRAN course about three hours long, given all in one evening, enabling me to transition to an IBM 704
from Andy's Olivetti calculator with its paper tape and the Chemistry Department's hand-cranked Marchant calculators.
Programming was in the ``MAD'' language. Michigan's Algorithmic Decoder language honored the magazine fixture
Alfred E. Neuman.  I also had the good luck to hear George Uhlenbeck's physics lectures on kinetic theory, transcribed
to the board from a musty notebook he held at arm's length.

Chemistry students at Michigan typically prepared seminars based on recent Scientific American articles. I was
fortunate to read Berni Alder and Tom Wainwright's ``Molecular Motions'' article ( see {\bf Figure 1} ) in the October
1959 Scientific American\cite{b2}. That article piqued my interest in Livermore with the goal of doing molecular
dynamics.  But first came some useful additional study, a post-doctoral year with Jacques Poirier in Durham North
Carolina where I learned to evaluate cluster integrals by Monte Carlo integration.

\pagebreak 
\begin{figure}
\includegraphics[width=2.5in,angle=+90.]{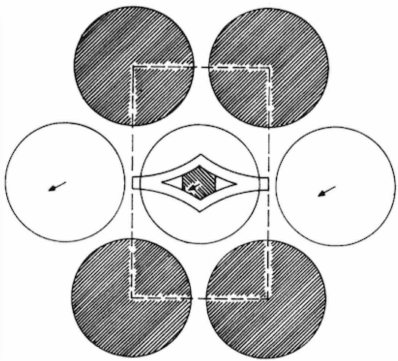}
\caption{
The diamond-shaped free volumes in the {\it correlated} cell model are larger than the usual hexagonal free volume
from the ordinary cell model, shown hatched at the center of the cell. When the second-neighbor separation exceeds
two diameters the white disks can escape their cells.  The pressure for the correlated model exhibits the van der
Waals'-like loop shown in Figure 3.
}
\end{figure}
\nopagebreak
\begin{figure}
\includegraphics[width=4.0in,angle=+90.]{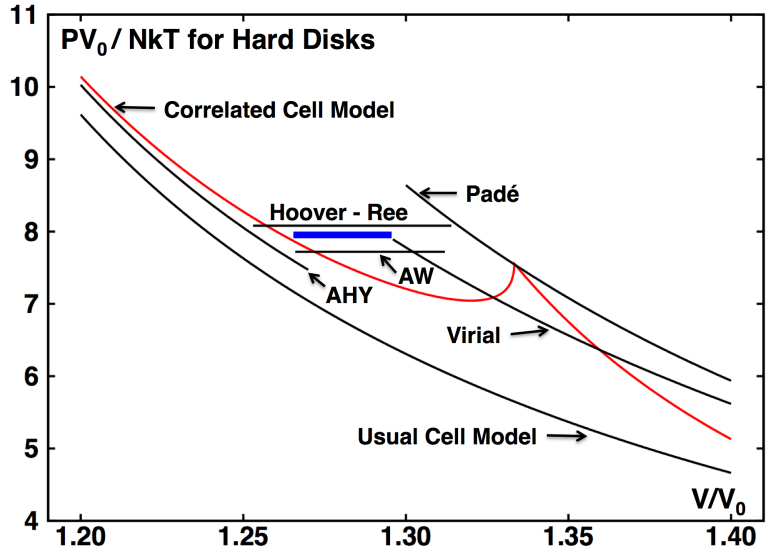}
\caption{
Hard-disk models. Pressure Volume plot for the correlated cell model is shown in red. $V_0$ is the close-packed
volume.  The estimated freezing pressures
from references 6 (AW) and 7 (Hoover-Ree) are based on molecular dynamics and a Pad\'e approximant respectively.
The high-density equation of state ( AHY ) is described in reference 8.  The most recent estimated pressure, in blue,
is taken from reference 4.  ``Virial'' is the ten-term expansion using Clisby and McCoy's virial coefficients from
reference 1.
}
\end{figure}
\newpage

\subsection{From Durham North Carolina to Livermore California}

\begin{figure}
\includegraphics[width=4.0in,angle=-90.]{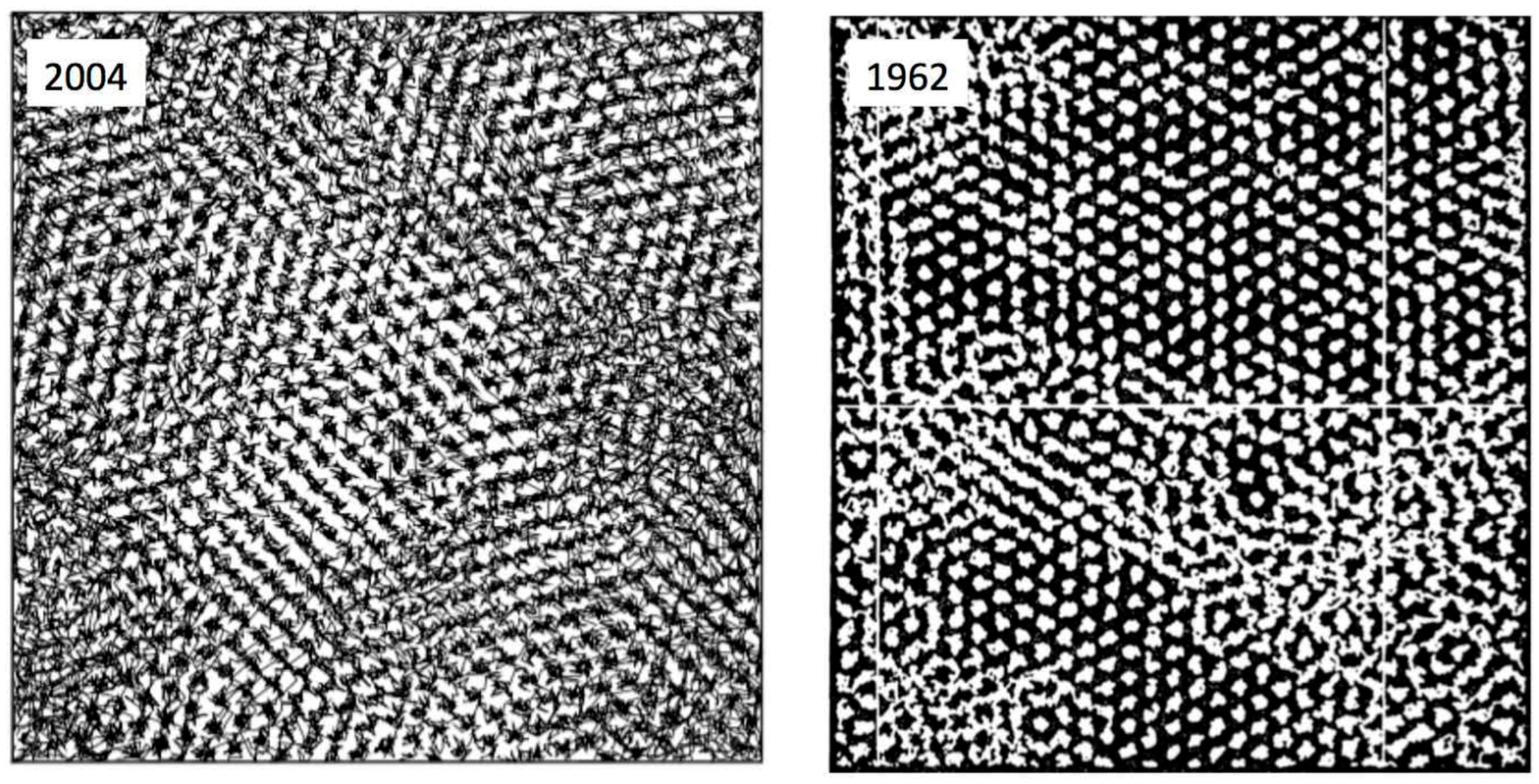}
\caption{
In the hexatic phase the particle motions from Zangi and Rice's 2004 simulation\cite{b5} parallel the three triangular
lattice directions.  Compare to the similar particle trajectories which can be found in the multiphase trajectory
taken from Alder and Wainwright's 1962 study\cite{b6}.
}
\end{figure}

By the time I left Michigan for a postdoc at Duke I was hooked on computer simulation. During the spring of 1962 I
interviewed at Livermore, with Berni and Tom, and at Los Alamos, with Bill Wood.  Livermore won out.  Besides the
better money for me it was much more interesting to see {\it motion} governed by differential equations rather than
watching uncorrelated Monte Carlo {\it moves}, no matter how clever the underlying algorithm.  Berni made it easy
for me to learn and to work with him and some of his many colleagues, Brad Holian, Francis Ree, Tom Wainwright, and
David Young.  We worked on a variety of projects in kinetic theory and statistical mechanics, mostly directed toward
equation of state properties for hard particles.

Our first project\cite{b3} ( see {\bf Figures 2 and 3} ) provided a great
example of advancing theory by the observation and description of computer experiments.  Movies of hard-disk dynamics
as well as Bill Wood's reports of his Monte Carlo studies at Los Alamos, revealed extensive cooperative motions of disks,
both linear and circular in nature. These correlated motions suggested modifying the cell model for dense fluids. That
model approximates the $N$th root of the partition function by a single-particle free-volume integral.  This good idea
is exact for a very light particle which moves so rapidly that its neighbors appear to be stationary. A ``correlated''
cell model based on the idea of correlated motion was my first Livermore publication.  The resulting van der
Waals'-like loop closely matched the large-system transition from fluid to solid as it was estimated in the 1960s.
A more precise idea of what happens at the melting and freezing densities was intensively investigated 50 years
later\cite{b4}.  Although the details involve a ``hexatic'' phase\cite{b5} ( see {\bf Figure 5} ) on the solid side of
the transition, from the visual standpoint the transition appears to be first-order.  The current estimate for the
transition pressure is just 3 percent higher than Berni and Tom's 1962 estimate\cite{b6} and 1.6 percent lower than our
best guesses with Francis Ree and Dave Young in 1968\cite{b7,b8} as shown in {\bf Figure 3}.

\begin{figure}
\includegraphics[width=4.0in,angle=+90.]{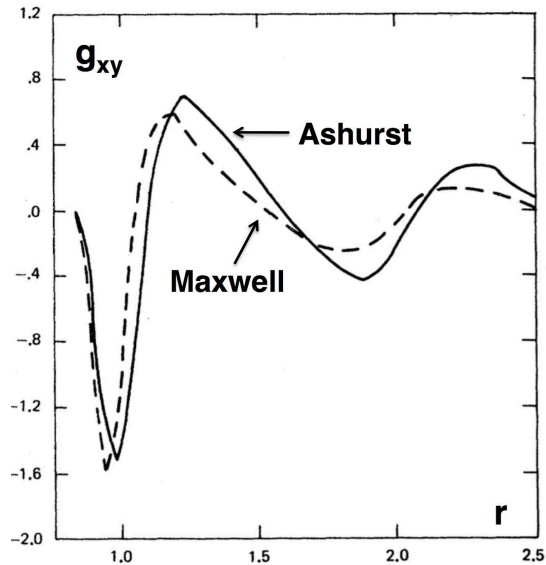}
\caption{
Nonequilibrium pair distribution function $g_{xy}$ which contributes to the potential shear stress ( from Ashurst's
Thesis\cite{b11} ).
Measurement with molecular dynamics is the solid line.  The dashed line corresponds to the rigid shear of the
equilibrium distribution function for a best-fit choice of Maxwell's relaxation time. The actual and approximated
contributions differ by about four percent.
}
\end{figure}

\subsection {Gordon Conferences in New Hampshire and CECAM workshops in France}

With the equilibrium equation-of-state problem solved by a van-der-Waals'-like perturbation theory\cite{b9} it was
time to move on. By 1971 Berni had helped me into a teaching position at Teller Tech, where my first PhD student,
Bill Ashurst, from just across the street at the Sandia Laboratory, shared my interest in {\it nonequilibrium}
molecular dynamics.  We developed numerical methods for measuring transport coefficients by modifying Hamiltonian
mechanics to include velocity and temperature gradients.  That work led to participating in Gordon Conferences in New
Hampshire, meeting scientists from all over the world, and to travel abroad.  I participated in Carl Moser's CECAM
workshops, originally in Orsay and later at Lyon, getting acquainted with many of the men interested in molecular
dynamics and visiting their laboratories: Giovanni Ciccotti, Gianni Iacucci, Dimitri Kusnezov, Michel Mareschal,
Shuichi Nos\'e, Loup Verlet, Bob Watts, Kris Wojciechowski, and Les Woodcock. In those days it was exciting and
interesting to establish the agreement between the Monte Carlo and molecular dynamics methods for systems as small
as four particles\cite{b10}. Molecular dynamics had the advantages of simulating actual motions while solving nonequilibrium
problems and generating nonequilibrium distribution functions\cite{b11}. See Ashurst's {\bf Figure 5}.

\subsection{Sabbatical in Canberra Australia, 1977-1978}

\begin{figure}
\includegraphics[width=4.5in,angle=+90.]{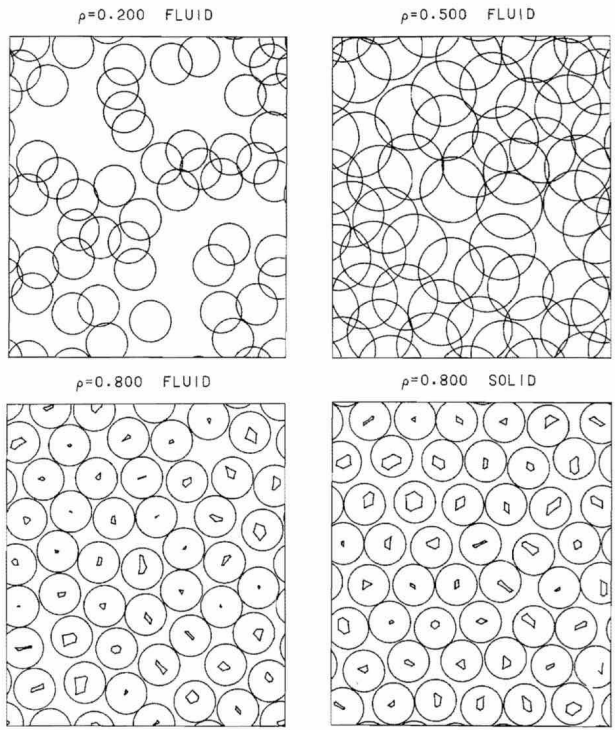}
\caption{The fluid configurations at the top illustrate extensive ( left ) and intensive ( right ) free volumes for hard
disks with half the diameter of the exclusion disks shown in the figure.  The ``percolation transition'' separating
the two regimes is at a density one fourth of the close-packed density.  The fluid and solid configurations at the
bottom show that the fluid-phase free volume is considerably smaller than the solid-phase free volume at the same density.
}
\end{figure}

\begin{figure}
\includegraphics[width=5.0in,angle=-90.]{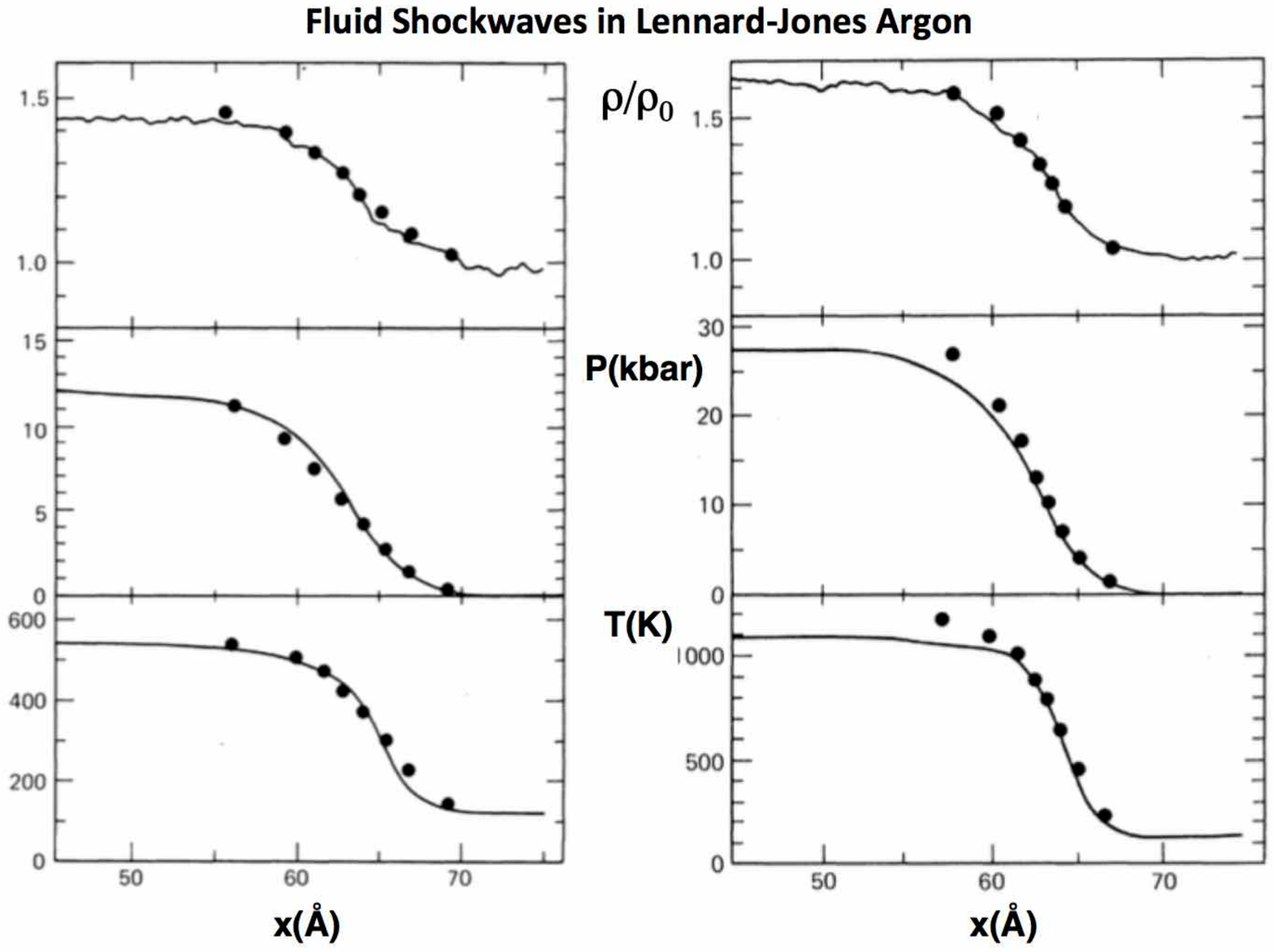}
\caption{
Relative density, pressure in kilobars, and temperature in kelvins for shock compressions of liquid argon to 12  and
to 27 kilobars.  My Navier-Stokes estimates ( filled circles ) from reference 15 are compared to Klimenko and Dremin's
simulation data ( lines ) from reference 14.
}
\end{figure}

Bill Ashurst had finished his thesis work, simulating nonequilibrium shear and heat flows with pair potentials,
in 1974.  I arranged a Fulbright leave to extend the pair-potential calculations to water with Bob Watts at
Canberra.  Once I had arrived two setbacks disabled the plan: [ 1 ] Bob was appointed head of the Computer Centre
within a week of my arrival and [ 2 ] his potential for water turned out to be unstable. As a result I mainly
worked with my son Nathan, evaluating hard-disk free volumes in the solid and fluid phases and locating the
percolation transition where the free volume changes from intensive to extensive\cite{b12}.  Two ways of illustrating
the free volumes ( disks of radius $\sigma$ above and diameter $\sigma$ below ) appear in {\bf Figure 6}.

Our office was just across the hall from a door with a hand-lettered sign identifying the occupant, a recent PhD and
enthusiastic bush-walker, ``BigFoot Evans''. Along with the Watts', we spent a lot of time together socially and in
the mountains, despite Bob's immersion, if not quite drowning, in administrative work. Almost every weekend found us
exploring the Australian bush or rock climbing at Booroomba Rocks, just half an hour outside Canberra.

In 1967 I set out to model the propagation of strong shockwaves in solids, strong enough to cause melting. Although a
preliminary account was published\cite{b13} the project was never completed due to the unreliability of the magnetic
tapes on which the particle coordinates and velocities were stored.  Large-scale shockwave simulations were put on hold
until around 1980.  By that time Klimenko and Dremin had published shockwave profiles\cite{b14} for two different shock
strengths. Their results are the solid lines in {\bf Figure 7}. In 1979 I compared their molecular dynamics simulations
of shockwaves to the predictions of Navier-Stokes continuum mechanics\cite{b15}. The good agreement at 12
and 27 kilobars set the stage for a large-scale higher-pressure effort at nearly 400 kilobars.  The work involved seven
of us confronting computer simulations of transport coefficients with the high-pressure shockwave data\cite{b16,b17}.
The agreement was semiquantitative. The observed viscosity at 400 kilobars was about thirty percent higher than
the low-strainrate Newtonian viscosity. Steady shear experiments predict a decrease rather than an increase.  Thus there
is still some interesting work to be done in order to understand this difference in the rate dependence.

\subsection{1984: Shuichi Nos\'e's Canadian Ideas Appear in France}

After his thesis work in Kyoto Shuichi Nos\'e moved to faroff Northern Canada to a postdoctoral position with Mike Klein.
Nos\'e  published two amazing papers in 1984 showing how to do constant {\it temperature} molecular dynamics in the
canonical ensemble\cite{b18,b19}. He used a logarithmic potential to store the excess energy from fluctuations around the
mean.  Their importance was clear.  He was invited to a CECAM workshop at Orsay, just outside Paris, to discuss his 
work.  After reading his papers I arranged to attend that CECAM workshop.  On the way there I came across Nos\'e on
an Orly-to-Paris train platform, a few days before the workshop was to begin. This was very good luck ! Back in
Livermore I had prepared a list of a dozen discussion questions for him concerning his papers. Most of them had
to do with ``time-scaling'', an important step in his work and one that was entirely foreign to me.  The timescaling
variable $s$ varies with time between zero and unity. $s$ appears in the denominator of the kinetic energy in
Nos\'e's Hamiltonian so that the equation of motion for the friction coefficient
$\dot \zeta \equiv \dot p_s = -(\partial{\cal H}/\partial s)$, for the motion of the momentum conjugate to $s$, with $s^3$ in the
denominator, is quite stiff :
$$
{\cal H}_{Nos\acute{e}} = (K/s^2) + \Phi + (\zeta^2/2) + NkT\ln(s) \ ;
\ K = \sum(p^2/2m) \ ; \ \Phi = \sum_{pairs} \phi \ .
$$

Nos\'e and I went over the new concepts carefully on a bench in front of the Notre Dame cathedral. That informative
meeting, together with the stimulation from the workshop that followed led me to spend the next two weeks after the
workshop at Philippe Choquard's laboratory in Lausanne. One of his students helped me to make Tektronix plots of
harmonic oscillator canonical trajectories, shown in {\bf Figure 8}. In the student's words ``we make a graphique'' !
It was amazing to me that the $\{ \ q,p,s,\zeta \ \}$ trajectories generated by two entirely different sets of
differential equations, one stiff and the other not, were identical.  The original stiff equations were  relatively
useless. The ``scaled'' equations with all of the rates multiplied by $s$ , were well-behaved and quite useful in
equilibrium canonical-ensemble simulations.

\begin{figure}
\includegraphics[width=3.5in,angle=+90.]{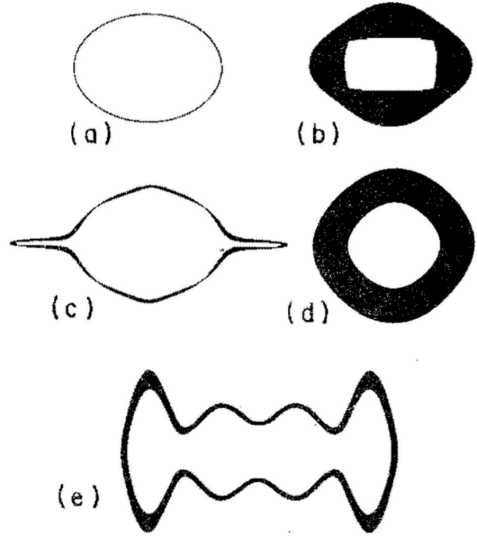}
\caption{
Tektronix plots of thermostated harmonic oscillator trajectories from reference 21. (a) is the ordinary isoenergetic
harmonic oscillator. (b) and (d) are thermostated oscillator trajectories while in (c) and (e) the thermostat response
$\dot \zeta$ has been increased by a factor of ten relative to the equations in the text.
}
\end{figure}

To illustrate, Nos\'e's original and ``scaled'' equations of motion for a harmonic oscillator are as follows :
$$
\{ \ \dot q = (p/s^2) \ ; \ \dot p = - q \ ; \ \dot s = \zeta \ ;
\ \dot \zeta = (p^2/s^3) - (T/s) \ \} \ [ \ {\rm Nos\acute{e}} \ ] \ ;
$$
$$
\{ \ \dot q = (p/s) \ ; \ \dot p = - qs \ ; \ \dot s = \zeta s \ ;
\ \dot \zeta = (p^2/s^2) - T \ \} \ [ \ {\rm scaled} \ ] \ .
$$
Nos\'e's four scaled first-order equations can then be simplified, eliminating both $p$ and $s$ to give a second-order
coordinate-space ``thermostated'' equation of motion along with a first-order ``feedback'' equation for the friction
coefficient $\zeta$ :
$$
\{ \ \ddot q = -q -\zeta \dot q \ ; \ \dot \zeta = \dot q^2 - T \ \} \ [ \ {\rm Nos\acute{e}-Hoover} \ ] \ .
$$
$\zeta $, originally the momentum associated with $s$, now plays the role of a ``friction coefficient'', or ``thermostat
variable''. The missing equation, $\dot s = s\zeta$ , is best ignored, as $s$ plays no role in the time-development
of the oscillator trajectory. In the simpler ``Nos\'e-Hoover'' approach the time derivative of $\zeta$ is derived
directly from Liouville's flow equation rather than from Nos\'e's ``{\it ad hoc}'' Hamiltonian.

It wasn't until twelve years later, 1996, that Dettmann and Morriss published a method\cite{b20} that avoids the explicit
time-scaling step by the equivalent two-part process of [ 1 ] multiplying the Hamiltonian by $s$ and [ 2 ] setting
the value of the Hamiltonian equal to zero.  Though unconventional it turns out that this same idea actually works
in general if one wishes to multiply the rates by an {\it arbitrary} function provided that the Hamiltonians are
also set equal to zero :
$$
{\cal H}_{DM} \equiv s{\cal H}_{Nos\acute{e}} \longrightarrow \{ \dot q,\dot p,\dot \zeta \ \}_{DM} =
s\{ \dot q,\dot p,\dot \zeta \ \}_{Nos\acute{e}} \ ; \ \dot \zeta_{DM} = -{\cal H}_{Nos\acute{e}} +
s\dot \zeta_{Nos\acute{e}} = s\dot \zeta_{Nos\acute{e}} \ . 
$$

In Lausanne after the Orsay meeting, and with a good understanding of Nos\'e's work, I wrote a short paper
describing my own interpretation of his work along with a simpler derivation of his results.  My Liouville equation
approach avoided Nos\'e's time-scaling. I included applications to the harmonic oscillator
and the isobaric ensemble.  Those three pages\cite{b21}, with the Lausanne Tektronix plots, describing what is now
called ``Nos\'e-Hoover'' mechanics have generated many extensions in directions well beyond my understanding. Oddly
enough I haven't since used this scaling trick with any other problem, but it looks like a good one to remember
nonetheless.
 
The microcanonical oscillator is the simplest of problems. It gives a circular phase-plane orbit. {\bf Figure 9}
illustrates examples of the complexity of the Nos\'e-Hoover oscillator dynamics in the canonical ensemble. Six
percent of the stationary distribution function for the oscillator's phase space ,
$$
f_{\rm stationary} = (2\pi)^{-3/2}e^{-q^2/2}e^{-p^2/2}e^{-\zeta^2/2} \ ,
$$
is occupied by a connected chaotic sea made up of solutions which separate exponentially fast from one another\cite{b22}.
The remaining 94\% of the distribution is made up of tori surrounding stable periodic orbits.  So this oscillator
model is far from ergodic.

\begin{figure}
\includegraphics[width=3.5in,angle=-90.]{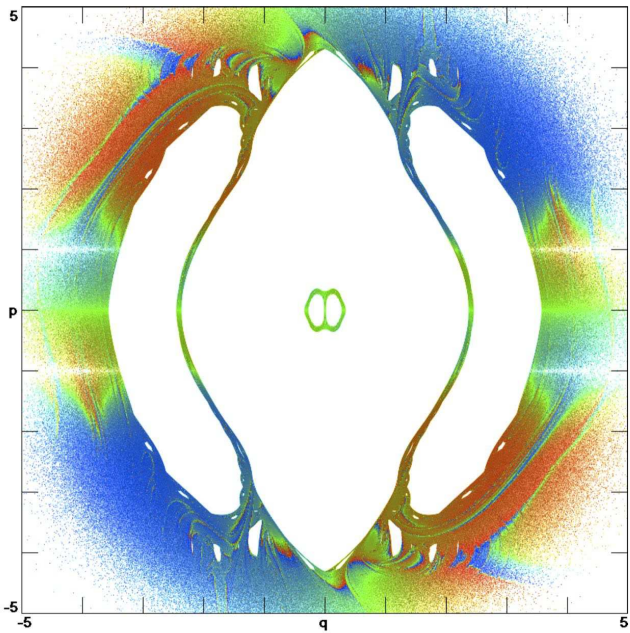}
\caption{Flux through the $\zeta=0$ cross section for the Nos\'e-Hoover oscillator showing $p(q)$ at each
crossing.  The white ``nullclines'' show that the flux vanishes for $p=\pm 1$ . All these points are located in the
chaotic sea, which makes up approximately six percent of the oscillator's stationary-state measure in
$(q,p,\zeta)$ phase space. Red/blue indicate the most positive/negative $\lambda_1(t)$ .
}
\end{figure}

Several ways of solving the harmonic oscillator problem with an ergodic dynamics were developed over the next
thirty years, all of them requiring at least two thermostat variables so as to produce chaos everywhere in
the oscillator phase space. An example from 1996\cite{b23} is
$$
\{ \ \dot q = p \ ; \ \dot p = -q - \zeta p - \xi p^3 \ ;
 \ \dot \zeta = p^2 - 1 \ ; \ \dot \xi = p^4 - 3p^2 \ \} \ .
$$
The two friction coefficients $\zeta$ and $\xi$ control the two moments $\langle \ p^2 \ \rangle$ and
$\langle \ p^4 \ \rangle$, respectively.  The four differential equations have a stationary and ergodic
phase-space distribution, Gaussian in all four variables :
$$
f_{\rm stationary} = (2\pi)^{-2}e^{-q^2/2}e^{-p^2/2}e^{-\zeta^2/2}e^{-\xi^2/2} \ ,
$$

\begin{figure}
\includegraphics[width=3.5in,angle=-90.]{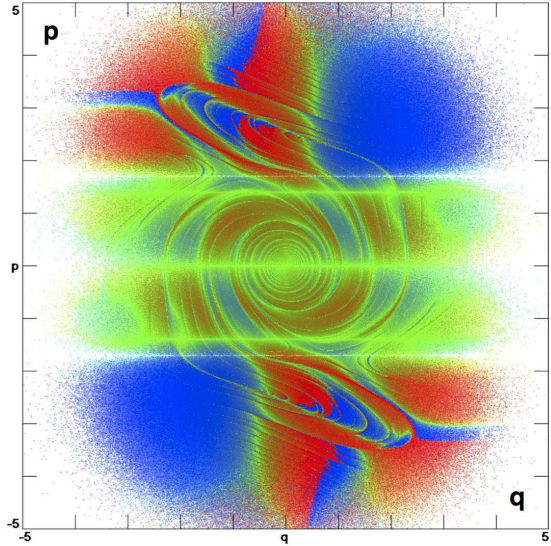}
\caption{
Flux through the $\zeta = 0$ cross-section for the 0532 oscillator reflects the ergodic nature of the model.  
The largest-to-smallest local exponents $\lambda_1(t)$ are colored from red-to-blue.  The local Lyapunov
exponents give the instantaneous rate of separation, $(\dot \delta/\delta)$ , for two nearby trajectories.
Notice $\lambda(q,p) \neq -\lambda(q,-p)$ while $\lambda(q,p) = \lambda(-q,-p)$ , showing a surprising symmetry
breaking !
}
\end{figure}

Thirty years after the original oscillator work, in the summer of 2015, I was spending a day watering trees at
our home in Ruby Valley. I had the idea of applying ``weak control'' to the harmonic oscillator, using a single
thermostat variable for the simultaneous control of {\it two} oscillator moments rather than just one. Successful
examples soon followed.  One of them, shown in {\bf Figure 10} is the ``0532 Model'' :
$$
\{ \ \dot q = p \ ; \ \dot p = -q -\zeta(0.05p + 0.32p^3) \ ;
 \ \dot \zeta = 0.05(p^2 - 1) + 0.32(p^4 - 3p^2) \ \} \ .
$$
Because this model {\it is} ergodic\cite{b24} its cross sections have no holes and are simple Gaussian functions.
But they take on some interest when colored according to the local values of the Lyapunov exponent. Despite
the time reversibility of the equations of motion the time-reversed trajectory's local
( instantaneous ) values of the largest Lyapunov exponent are unrelated to the largest local exponent going
forward in time.  This symmetry breaking is a conservative relative of the dissipative symmetry breaking
which generates ``strange attractors'' ( and not ``strange repellors'' ) in nonequilibrium simulations.  With
the successful application of weak control the original 1984 problem of finding an ergodic
thermostat for the harmonic oscillator finally produced several single-thermostat solutions.

Continued exploration of the oscillator phase space is ongoing. The most recent work I have seen is a collection of
{\it knots}, tied in phase space by one or two Nos\'e-Hoover oscillator trajectories\cite{b25}.  Piotr Pieranski's internet
descriptions and depictions of knots are an enriching experience.  Knot theory was and is a well-developed field of
study going back to Kelvin and Maxwell.  For an introductory look at knots see the interesting work on the prototypical overhand
( or trefoil ) knot\cite{b26}.  See {\bf Figure 11}.

\subsection{1985: Sabbatical Leave at Boltzmanngasse in Wien}

The CECAM meetings enriched my life, leading to a sabbatical with Harald Posch and Karl Kratky in Vienna and
another four years later with Shuichi Nos\'e in Yokohama.  Work at Harald's Boltzmanngasse laboratory was intense
and productive.  Apart from the floor and the high ceiling Harald and Franz Vesely and I covered all of the horizontal
surfaces of Harald's large laser lab with pictures and projections of tori and cross-sections of chaotic seas\cite{b27}.
The cross-section calculations were time-consuming, taking all night to run, showing up on the screen in the
morning in what I recall was a beautiful shade of green. To assess the improvement in computation I recalculated
a cross-section problem on a Mac laptop, generating two billion timesteps solving the three differential equations
describing the simplest Nos\'e-Hoover thermostated oscillator :
$$
\{ \ \dot q = p \ ; \ \dot p = -q - \zeta p \ ; \ \dot \zeta = p^2 - 1 \ \} \ .
$$
It took exactly two minutes to run, generating about half a billion points for each of the three cross sections :
$\{ \ q=0, \ p=0, \ \zeta=0 \ \}$.  Comparing the 2016 and 1985 computation times gave a speedup of $500 \simeq 2^9$
or so in 30 years, somewhat slower than Moore's Law's prediction, but a very pleasant outcome of the computer
revolution.

\begin{figure}
\includegraphics[width=5.5in,angle=-90.]{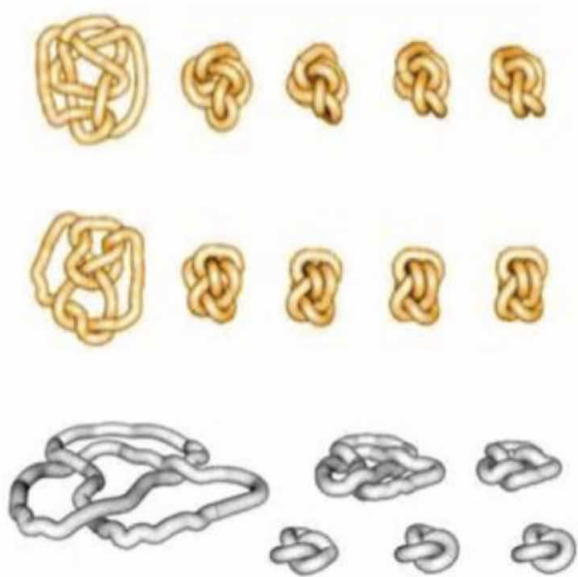}
\caption{
Simple examples of Piotr Pieranski's software for tightening and rendering knots.  The trefoil ( or overhand ) 
knot below can be found among the many Nos\'e-Hoover oscillator orbits explored in reference 25.
}
\end{figure}

By 1987 Bill and another excellent PhD student Bill Moran, had analyzed the isokinetic motion of a point mass
falling through a triangular lattice of scatterers.  This is the Galton Board\cite{b28}, named for the statistician who
built one over 100 years ago, to demonstrate the binomial and Gaussian distributions caused by scattering. In
between collisions the Galton-Board equations of motion are ;
$$
\dot x = p_x \ ; \ \dot y = p_y \ ; \ \dot p_x = -\zeta p_x \ ; \ \dot p_y = - E - \zeta p_y \ .
$$
The friction coefficient $\zeta = -Ep_y$ maintains the squared velocity equal to unity, $p_x^2 + p_y^2 \equiv 1$ ,
as the reader can easily verify.  The dissipation induced by the friction coefficient $\zeta$ allows the moving
particle to descend, converting field energy to heat which is extracted by the reservoir.  A look at the phase-space
cross sections in {\bf Figure 12} shows that the dimensionality of the fractal attractor decreases as the field
strength is increased from one, to two, to three, to four.  In this last case the phase space is separated into tori,
describing the stable bouncing of the moving particle between two scatterers in the same horizontal row, as well as
the chaotic sea which describes the dissipation associated with the Second Law of Thermodynamics.  This problem
with its motion on a three-dimensional energy surface in the four-dimensional phase space, was one of the earliest
to show the multifractal phase-space distributions characterizing time-reversible deterministic dissipative
systems.

\begin{figure}
\includegraphics[width=5in,angle=-90.]{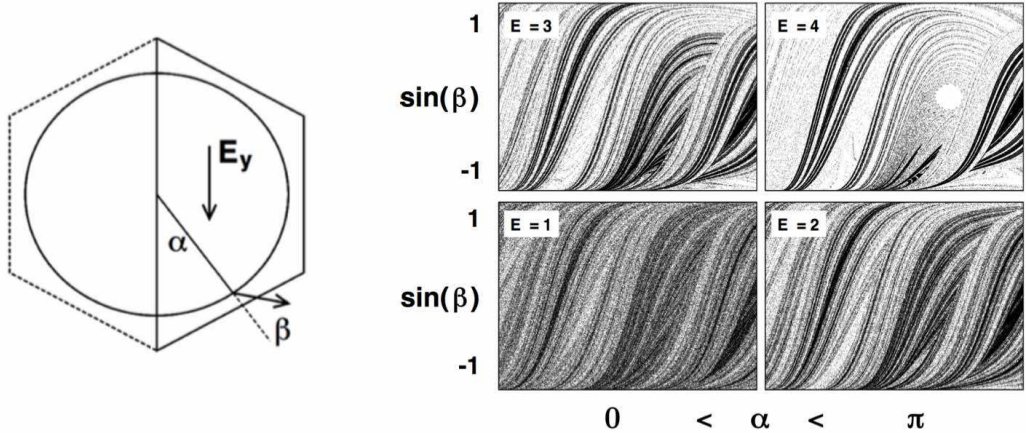}
\caption{
A unit cell of the triangular lattice Galton Board\cite{b28} is shown to the left with the definitions of the two
angles $\alpha$ and $\beta$ defining a scatterer collision.  At the right phase-space cross sections are shown for
field strengths of 1, 2, 3, and 4.  In the last case there are concentric tori occupying the near-circular hole
in the section.  At the lower field strengths the dynamics is ergodic so that all $(\alpha,\beta)$ states are a
part of the chaotic sea.
}
\end{figure}

\subsection{Moral}
I am very grateful to all my colleagues for support, wisdom, collaborations, friendship, and love over the years.
I am particularly grateful to Berni Alder for his extending a helping hand in so many ways that proved crucial to
the good life I have enjoyed.  His generosity in presenting an inspiring talk at the Sheffield Conference was a
very welcome eightieth Birthday gift. I urge those of you who are younger to reflect upon your good fortune in
being a part of our progress in understanding the world around us.  

\pagebreak

\end{document}